\documentclass[aps,twocolumn,a4paper,pra,superscriptaddress,floatfix]{revtex4}
\usepackage{amssymb}
\usepackage{amsmath}
\usepackage{amsfonts}
\usepackage{graphicx}
\def\P{\mathrm{P}}
\def\C{\mathrm{C}}

\begin{document}
\title{Reply to ``Comment on ``Lateral Casimir Force beyond the Proximity Force Approximation'' ''}
\author{Robson B. Rodrigues}
\author{Paulo A. Maia Neto}
\affiliation{Instituto de F\'{\i}sica, Universidade Federal do Rio de Janeiro, 
Caixa Postal 68528, Rio de Janeiro, RJ, 21941-972, Brazil}
\author{Astrid Lambrecht}
\author{Serge Reynaud}
\affiliation{Laboratoire Kastler Brossel,
CNRS, ENS, UPMC case 74, Campus Jussieu, 75252 Paris, France}
\date{\today}
\maketitle

Our letter \cite{letter} is devoted to the presentation of a
novel theoretical approach to the lateral Casimir force beyond
the regime of validity of the ``Proximity Force Approximation'' (PFA). 
The approach relies on scattering theory used in a perturbative expansion 
\cite{scat_pert} valid when the corrugation amplitudes $a_1,a_2$ are smaller 
than the 3 other length scales, the mean separation distance $L$, the 
corrugation period $\lambda _\C$ and the plasma wavelength $\lambda _\P$. 
This restriction is repeatedly stressed in the abstract and the body text 
of \cite{letter} and it is also the main topic of the comment \cite{comment}.
We agree with the statements in the comment which constitute yet another 
warning that the calculations presented in \cite{letter} are valid 
``provided that the corrugation amplitude is smaller than the other length scales'' 
(last sentence of \cite{letter}). 
But we strongly disagree with the idea that the approach in Ref.~\cite{letter} 
is not appropriate for making statements on the accuracy of the PFA.

It was natural to illustrate the results of the new 
approach by applying them to a comparison with the
experiment reported in Refs.~\cite{exp}. 
As the corrugation amplitudes in the experiment are smaller, but not much smaller, 
than the other length scales, the comparison could unfortunately
not be direct, as explained as fairly as possible in \cite{letter}.
The results of \cite{letter} are however of clear interest 
for the experiment, as they can be summed up as follows,
assuming that $L,\lambda_\C,\lambda_\P$ are chosen in accordance with
the experimental numbers~\cite{exp}~:  
i) the perturbative calculation beyond the PFA \cite{letter} gives a force 
approximately 40\% smaller than the perturbative calculation within the PFA;
ii) as the calculation of Refs.~\cite{exp} takes into account 
higher order powers in $a_1a_2$ (which is easy within the PFA), 
we extracted the perturbative result (proportional to $a_1a_2$) by 
discarding the higher orders contribution; this procedure produced 
a discrepancy of approximately 30\% between the two methods.
This number points to a potential concern for 
theory-experiment comparison, which is nevertheless not so severe 
as the experimental results (0.32$\pm $0.077pN according to
\cite{comment}) correspond to a relative accuracy of $\pm 24\%$. 

The focus of the comment \cite{comment} is an argument about our 
estimation of the discrepancy. 
The comment sidesteps the issue by comparing two numbers 
which are {\it not} to be compared (and which we did {\it not} compare),
namely the perturbative result beyond the PFA and the non perturbative 
result within the PFA.
It thus fabricates a large discrepancy (nearly 60\%) which would make the
concern more severe. We certainly do not approve this way of comparison 
since there is not any reason to ignore the effect of higher order 
corrections in one calculation and take it into account in the other one.
More work is needed in order to settle the issue of 
theory-experiment comparison.

Progress on this question could be achieved by calculating higher order 
corrections for metallic mirrors beyond the PFA.
These corrections are expected to affect the numbers, but
they will hardly compensate exactly the deviation 
from the PFA demonstrated in the perturbative regime. 
Let us underline at this point that the second paragraph
of the comment \cite{comment}, aimed at raising doubts on the
predictions of \cite{letter}, is based on a mistake.
The factor $\rho$, which measures the deviation from the PFA, is 
a function of the {\it three} length scales $L,\lambda _\C,\lambda _\P$, 
which is calculated in \cite{letter} for metallic mirrors.
The case of perfectly reflecting mirrors is recovered in the limit
$\lambda _\P\rightarrow 0$ but, 
in contrast with what is stated in \cite{comment}, 
the general function cannot be reconstructed from this particular limit.

Progress could alternatively come for experiments using smaller 
corrugation amplitudes while showing a better experimental accuracy.
The first condition would aim at reaching the condition 
$a_1,a_2\ll L,\lambda _\C,\lambda _\P$ which delineates the range of validity
of the theoretical predictions of \cite{letter}.
As emphasized in the conclusion of our letter, 
this would make possible ``an accurate comparison between theory
and experiment in a configuration where geometry plays a 
non trivial role, {\it i.e.} beyond the PFA.''
Meanwhile, an improved accuracy would allow to compare experiment with
different theoretical predictions.


\begin{thebibliography}{9}
\bibitem{letter} R.B. Rodrigues, P.A. Maia Neto, A. Lambrecht and S.
Reynaud, Phys. Rev. Lett. \textbf{96}, 100402 (2006).

\bibitem{scat_pert} P.A. Maia Neto, A. Lambrecht and S. Reynaud, 
Europhys. Lett. \textbf{69}, 924 (2005);
Phys. Rev. \textbf{A 72}, 012115 (2005).

\bibitem{comment} F. Chen, U. Mohideen, G.L. Klimchtskaya and V.M.
Mostepanenko, ``Comment on \cite{letter}'', submitted to Phys. Rev. Lett.  

\bibitem{exp} F. Chen, U. Mohideen, G.L. Klimchtskaya and V.M.
Mostepanenko, Phys. Rev. Lett. \textbf{88}, 101801 (2002);
Phys. Rev. \textbf{A 66}, 032113 (2002).

\end{thebibliography}
\end{document}